\documentclass[%
 reprint,
superscriptaddress,
 amsmath,amssymb,
 aps,
floatfix,
]{revtex4-1}

\usepackage{graphicx}% Include figure files
\usepackage{dcolumn}% Align table columns on decimal point
\usepackage{bm}% bold math
\usepackage{braket}
\usepackage{multirow}
\usepackage{dsfont}
\usepackage{mathtools}
\usepackage{xcolor}
\usepackage[normalem]{ulem}

\begin{document}

\preprint{KEK-CP-360}
\preprint{RBRC 1243}

\title{Approximate degeneracy of $J=1$ spatial correlators in high temperature QCD}

\author{C.~Rohrhofer}
\affiliation{Institute of Physics, University of Graz, 8010 Graz, Austria}
\author{Y.~Aoki}
\affiliation{KEK Theory Center, High Energy Accelerator Research Organization (KEK), Tsukuba 305-0801, Japan}
\affiliation{RIKEN BNL Research Center, Brookhaven National Laboratory, Upton NY 11973, USA}
\author{G.~Cossu}
\affiliation{School of Physics and Astronomy, The University of Edinburgh, Edinburgh EH9 3JZ, United Kingdom}
\author{H.~Fukaya}
\affiliation{Department of Physics, Osaka University, Toyonaka 560-0043, Japan}
\author{L.~Ya.~Glozman}
\affiliation{Institute of Physics, University of Graz, 8010 Graz, Austria}
\author{S.~Hashimoto}
\affiliation{KEK Theory Center, High Energy Accelerator Research Organization (KEK), Tsukuba 305-0801, Japan}
\affiliation{School of High Energy Accelerator Science, The Graduate University for Advanced Studies (Sokendai), Tsukuba 305-0801, Japan}
\author{C.~B.~Lang}
\affiliation{Institute of Physics, University of Graz, 8010 Graz, Austria}
\author{S.~Prelovsek}
\affiliation{Faculty of Mathematics and Physics, University of Ljubljana, 1000 Ljubljana, Slovenia}
\affiliation{Jozef Stefan Institute, 1000 Ljubljana, Slovenia}
\affiliation{Institute f\"ur Theoretische Physik, Universit\"at Regensburg, D-93040, Germany}
 
\date{\today}

\begin{abstract}
We study spatial isovector meson correlators in $N_f=2$ QCD with dynamical
domain-wall fermions on $32^3\times 8$ lattices at temperatures $T=220-380$~MeV.
We measure the correlators of spin-one ($J=1$) operators including vector,
axial-vector, tensor and axial-tensor. Restoration of chiral $U(1)_A$  and
$SU(2)_L \times SU(2)_R$ symmetries of QCD implies  degeneracies in
vector--axial-vector ($SU(2)_L \times SU(2)_R$) and tensor--axial-tensor ($U(1)_A$)
pairs,  which are indeed observed at temperatures above $T_c$. Moreover, we 
observe an approximate degeneracy of all $J=1$ correlators with increasing
temperature. This approximate degeneracy suggests emergent $SU(2)_{CS}$ and
$SU(4)$ symmetries at high temperatures, that mix left- and right-handed quarks. 
\end{abstract}
%\pacs{Valid PACS appear here}
\maketitle

%\tableofcontents
%%%%%%%%%%%%%%%%%%%%%%%%%%%%%%%%%%%%%%%%%%%%%%%%%%%%%%%%%%%%%%%%%%%%%%%%%%%%%%
%%%%%%%%%%%%%%%%%%%%%%%%%%%%%%%%%%%%%%%%%%%%%%%%%%%%%%%%%%%%%%%%%%%%%%%%%%%%%%
%%%%%%%%%%%%%%%%%%%%%%%%%%%%%%%%%%%%%%%%%%%%%%%%%%%%%%%%%%%%%%%%%%%%%%%%%%%%%%
\section{\label{sec:intro}Introduction}

One of the key questions of QCD, that is of crucial importance both for
astrophysics and cosmology, is the nature of the strongly interacting matter
at high temperatures. This question attracts enormous experimental and
theoretical efforts. It is established in ab initio QCD calculations on the
lattice that there is a transition to the chirally symmetric regime where the
quark condensate, an order parameter of $SU(2)_L \times SU(2)_R$ chiral
symmetry, vanishes. At the same time there is strong evidence from
calculations with manifestly chirally-invariant lattice fermions that
above the critical temperature also the $U(1)_A$ symmetry gets restored and
a gap opens in the spectrum of the Dirac operator
\cite{Cossu:2013uua,Tomiya:2016jwr,Bazavov:2012qja}.

To get some new insight on the symmetry structure of high temperature QCD
we calculate spatial correlators of all possible spin $J=0$ and $J=1$ local
isovector operators using the chirally invariant domain-wall Dirac operator
in two-flavour QCD at different temperatures up to 380~MeV. All correlators
that are connected by $SU(2)_L \times SU(2)_R$ or $U(1)_A$  transformations
are the same within errors
at temperatures above the critical one, which is in
agreement with restoration of both symmetries at high temperature.
Surprisingly, we  also observe an approximate degeneracy of some correlators
that are connected neither by $SU(2)_L \times SU(2)_R$ nor by $U(1)_A$
transformations. 

The observed approximate degeneracies at high temperature are in agreement
with emergent $SU(2)_{CS}$ (chiral-spin) and $SU(4)$ symmetries
\cite{Glozman:2014mka, Glozman:2015qva}, 
which contain transformations that mix the left- and right-handed components
of quarks. These symmetries have been observed before in $T=0$ dynamical
calculations upon artificial truncation of the near-zero modes of the
overlap Dirac operator
\cite{Denissenya:2014poa,Denissenya:2014ywa,Denissenya:2015mqa,Denissenya:2015woa}.
The near-zero modes of the Dirac operator 
on the lattice
are strongly suppressed at high
temperatures
\cite{Cossu:2013uua,Tomiya:2016jwr},
which motivates our present exploration of the correlators and symmetries at
high temperatures, this time without truncating the Dirac eigenmodes.

%%%%%%%%%%%%%%%%%%%%%%%%%%%%%%%%%%%%%%%%%%%%%%%%%%%%%%%%%%%%%%%%%%%%%%%%%%%%%%
%%%%%%%%%%%%%%%%%%%%%%%%%%%%%%%%%%%%%%%%%%%%%%%%%%%%%%%%%%%%%%%%%%%%%%%%%%%%%%
%%%%%%%%%%%%%%%%%%%%%%%%%%%%%%%%%%%%%%%%%%%%%%%%%%%%%%%%%%%%%%%%%%%%%%%%%%%%%%
\section{\label{sec:sim}Simulation}

%%%%%%%%%%%%%%%%%%%%%%%%%%%%%%%%%%%%%%%%%%%%%%%%%%%%%%%%%%%%%%%%%%%%%%%%%%%%%%
%%%%%%%%%%%%%%%%%%%%%%%%%%%%%%%%%%%%%%%%%%%%%%%%%%%%%%%%%%%%%%%%%%%%%%%%%%%%%%
\subsection{\label{subsec:lattices}Lattices}

The gauge configurations used in the numerical simulation of QCD are generated
using the Symanzik gauge action and 
%$n_f=2$  M\"obius domain wall fermion action
two degenerate flavors of M\"obius domain wall fermions
\cite{Brower:2005qw,Brower:2012vk}.
The gauge links are stout smeared three times before the computation of the
Dirac operator.
The length in the fifth direction $L_s$ is chosen to achieve precise chiral
symmetry. The boundary conditions  for quarks are set antiperiodic in
$t-$direction, and periodic in spatial directions.
The ensembles and parameters including the lattice spacing $a$ are listed
in Table~\ref{tab:ensembles}.
The degenerate up and down quark masses $m_{ud}$ are set to 2--15~MeV, 
which is essentially negligible at the temperatures we studied, {\it i.e.}
$T\simeq$ 220--380~MeV.
More
details on the chiral properties for this set of parameters are given in
\cite{Cossu:2015kfa,Tomiya:2016jwr}.
We study spatial ($z$-direction) correlators as was first suggested in ref.
\cite{DeTar:1987xb} 
(see also
\footnote{The asymptotic slopes of z-correlators
were addressed  e.g. in  C. Bernard et al., (MILC Collaboration) Phys. Rev. Lett. 68 (1992) 2125;
E. Laermann and P. Schmidt, Eur. J. Phys. C 20 (2001) 541.hep-lat/0103037 ;
R. V. Gavai, S. Gupta and P. Majumdar, Phys. Rev. D 65 (2002) 054506.
 hep-lat/0110032 ;
S. Wissel et al, PoS LAT2005 (2006) 164;
hep-lat/0510031
E. Laermann et al (RBC-Bielefeld Coll.), PoS LAT2008 (2008) 193.
\cite{Gavai:2008yv,Cheng:2010fe,Banerjee:2011yd}.
}).

\begin{table}
\center
\begin{tabular}{c|c|c|c|c|c|c}
\hline\hline
$\beta$ & $m_{ud}a$ & $a$ [fm] & \# configs & $L_s$ & $T$ [MeV] & $T/T_c$ \\\hline
$4.10$  & 0.001 & $0.113$ &     800  & 24  & $\sim220$ & $\sim 1.2$ \\
$4.18$  & 0.001 & $0.096$ &     230  & 12  & $\sim260$ & $\sim 1.5$ \\
$4.30$  & 0.001 & $0.075$ &     260  & 12  & $\sim320$ & $\sim 1.8$ \\
$4.37$  & 0.005 & $0.065$ &     120  & 12  & $\sim380$ & $\sim 2.2$ \\
\hline\hline
\end{tabular}
% this table counts as 13+6.5 * 8 = 65 words
\caption{Gauge ensembles for $32^3 \times 8$ lattices used in this work.
$L_s$ denotes the length of the fifth dimension in the domain wall fermion
formulation. The critical temperature for this set of parameters
is $T_c=175\pm5$~MeV.}
\label{tab:ensembles}
\end{table}

%%%%%%%%%%%%%%%%%%%%%%%%%%%%%%%%%%%%%%%%%%%%%%%%%%%%%%%%%%%%%%%%%%%%%%%%%%%%%%
%%%%%%%%%%%%%%%%%%%%%%%%%%%%%%%%%%%%%%%%%%%%%%%%%%%%%%%%%%%%%%%%%%%%%%%%%%%%%%
\subsection{\label{subsec:ops}Operators}

The observables of interest are correlators of non-singlet local operators
$\mathcal{O}_\Gamma(x) = \bar q(x) \Gamma \frac{\vec{\tau}}{2} q(x)$,
where $\Gamma$ might be any combination of
$\gamma$-matrices, \textit{i.e.} the Clifford algebra, containing 16 elements;
$\tau_a$ are the isospin Pauli matrices.

A zero-momentum
projection is done by summation over all lattice points in slices orthogonal
to the measurement direction. When measuring in $z$-direction this means
\begin{equation}
C_\Gamma(n_z) = \sum\limits_{n_x, n_y, n_t}
\braket{\mathcal{O}_\Gamma(n_x,n_y,n_z,n_t)
\mathcal{O}_\Gamma(\mathbf{0},0)^\dagger}.
\label{eq:momentumprojection}
\end{equation}
For the Vector and Axial-vector operators $\Gamma$ has the
following components:
\begin{eqnarray}
\mathbf{V}=
\left(
\begin{array}{c}
\gamma_1 = V_x \\
\gamma_2 = V_y \\
\gamma_4 = V_t 
\end{array}
\right),
\quad
\mathbf{A}=
\left(
\begin{array}{c}
\gamma_1 \gamma_5 = A_x \\
\gamma_2 \gamma_5  = A_y \\
\gamma_4 \gamma_5  = A_t 
\end{array}
\right).
\label{eq:vectorcurrents}
\end{eqnarray}
Conservation of the vector current requires that $V_z$ does not propagate in
$z$-direction. 
As the axial vector current $j_5^\mu$ is not conserved at zero temperature,
the relevant component $\gamma_3 \gamma_5$ of the Axial-vector does
propagate at zero temperature and eventually couples to the
Pseudoscalar. Above the critical temperature --- after $U(1)_A$ and
$SU(2)_L \times SU(2)_R$ restoration ---
$A_z$ behaves as its parity partner $V_z$ and does not propagate in
$z$-direction.
For propagation in $z$-direction the tensor elements $\sigma_{\mu\nu}$
of the Clifford algebra are organized
in the following way in components of Tensor- and
Axial-tensor vectors:
\begin{eqnarray}
\mathbf{T}=
\left(
\begin{array}{c}
\gamma_1 \gamma_3 = T_x \\
\gamma_2 \gamma_3 = T_y \\
\gamma_4 \gamma_3 = T_t 
\end{array}
\right),
\;
\mathbf{X}=
\left(
\begin{array}{c}
\gamma_1 \gamma_3 \gamma_5 = X_x \\
\gamma_2 \gamma_3 \gamma_5 = X_y \\
\gamma_4 \gamma_3 \gamma_5 = X_t 
\end{array}
\right).
\label{tensorcurrents}
\end{eqnarray}
Table~\ref{tab:ops} summarizes our operators
and gives the  $U(1)_A$ and $SU(2)_L \times SU(2)_R$ relations of these operators. Given
restoration of the $U(1)_A$ and $SU(2)_L \times SU(2)_R$ symmetries at high $T$ we expect
degeneracies of correlators calculated with the corresponding operators.

\begin{table}
\center
\begin{tabular}{cccll}
\hline\hline
 Name        &
 Dirac structure &
 Abbreviation    &
 \multicolumn{2}{l}{
   %Symmetries
 } 
 \\ \hline
 %%%%%%%%%%%
\textit{Pseudoscalar}        & $\gamma_5$                 & $PS$         & \multirow{2}{12mm}{$\left.\begin{aligned}\\ \end{aligned}\right] U(1)_A$} &\\
\textit{Scalar}              & $\mathds{1}$               & $S$          & &\\\hline
\textit{Axial-vector}         & $\gamma_k\gamma_5$         & $\mathbf{A}$ & \multirow{2}{12mm}{$\left.\begin{aligned}\\ \end{aligned}\right] SU(2)_A$}&\\
\textit{Vector}                  & $\gamma_k$                         & $\mathbf{V}$ &        & \\
\textit{Tensor-vector}       & $\gamma_k\gamma_3$         & $\mathbf{T}$ & \multirow{2}{12mm}{$\left.\begin{aligned}\\ \end{aligned}\right] U(1)_A$} &\\
\textit{Axial-tensor-vector} & $\gamma_k\gamma_3\gamma_5$ & $\mathbf{X}$ &         &\\
\hline\hline
\end{tabular}
\caption{
Bilinear operators considered in this work and their transformation properties
(last column). This classification assumes propagation in $z$-direction. The
open vector index $k$ denotes the components $1,2,4$, \textit{i.e.} $x,y,t$.}
\label{tab:ops}
\end{table}

For measurements at zero temperature the three components of the vectors give
the same expectation value due to the $SO(3)$ symmetry in continuum.  
In our finite temperature setup this rotational symmetry is broken
and the residual $SO(2)$ symmetry in the $(x,y)-$plane
connects $V_x \leftrightarrow V_y, A_x \leftrightarrow A_y$
\textit{etc.} operators. 

On the lattice at finite temperature the symmetry is $D_{4h}$ where the vector components belong to one
two-dimensional ($V_x$,$V_y$) and one one-dimensional ($V_t$) irreducible
representations, and similar for $\mathbf{A}$, $\mathbf{T}$, $\mathbf{X}$. This is discussed in more detail
in Appendix A (see also \cite{Banerjee:2011yd}). Operators from different irreducible 
representations are not connected by the $D_{4h}$ transformations and consequently the $D_{4h}$ symmetry
of the lattice does not predict the $E_1,E_2,E_3$ multiplet structures discussed in Section III.
The $x$ and $y$ components of $\mathbf{V}$ have degenerate energy levels, and correspondingly
those for the other Dirac structures. We therefore show only $x$- and $t$-components
in the plots.
%%%%%%%%%%%%%%%%%%%%%%%%%%%%%%%%%%%%%%%%%%%%%%%%%%%%%%%%%%%%%%%%%%%%%%%%%%%%%%
%%%%%%%%%%%%%%%%%%%%%%%%%%%%%%%%%%%%%%%%%%%%%%%%%%%%%%%%%%%%%%%%%%%%%%%%%%%%%%
%%%%%%%%%%%%%%%%%%%%%%%%%%%%%%%%%%%%%%%%%%%%%%%%%%%%%%%%%%%%%%%%%%%%%%%%%%%%%%
\section{\label{sec:results}Results}

\begin{figure}
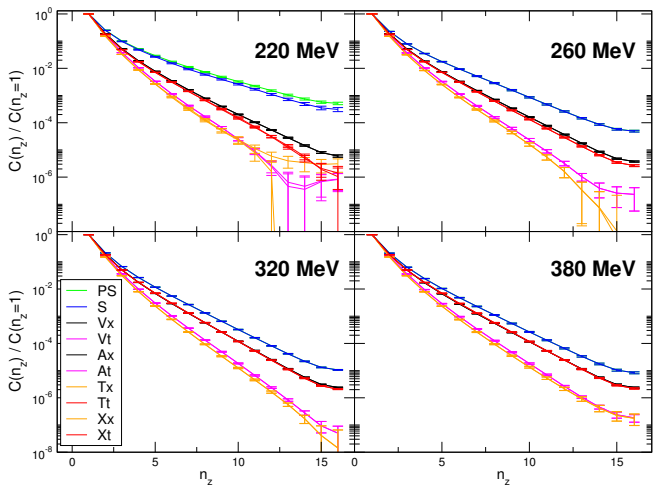

  \centering
  \includegraphics[scale=0.32]{{{1}}}
  \caption{
    Normalized spatial correlators. The temperatures correspond
    to the ensembles listed in Table~\protect{\ref{tab:ensembles}}.
  }
  \label{fig:corrs}
\end{figure}

Figure \ref{fig:corrs} shows the spatial correlation functions normalized to 1
at $n_z=1$ for the operators given in Table~\ref{tab:ops}.
As argument we show $n_z$ which is proportional to the dimensionless product
$zT$ for fixed $N_t$, the temporal extent of the lattice. 

As we describe in more detail later, we find that all correlators 
connected by $SU(2)_L \times SU(2)_R$ and $U(1)_A$ transformations coincide 
within small deviations at $T > 220$~MeV, which means that at these
temperatures both chiral symmetries get restored. 
More interestingly, there are additional degeneracies of correlators.
In total we observe three different multiplets:
\begin{eqnarray}
E_1: & \qquad PS \leftrightarrow S \label{eq:e1} \\
E_2: & \qquad V_x \leftrightarrow T_t \leftrightarrow X_t \leftrightarrow A_x \label{eq:e2} \\
E_3: & \qquad V_t \leftrightarrow T_x \leftrightarrow X_x \leftrightarrow A_t. \label{eq:e3}
\end{eqnarray}
$E_1$ describes the Pseudoscalar-Scalar multiplet connected
by the $U(1)_A$ symmetry,
that is realized in the absence of chiral 
zero-modes \cite{Kogut:1998rh}.
Note that we only consider the isospin triplet
channels so $S$ corresponds to the $a_0$- rather than the $\sigma$-particle.
The $E_2$ and $E_3$ multiplets on the other hand contain
some operators that are not connected by either
$SU(2)_L \times SU(2)_R$  or $U(1)_A$
transformations.

\begin{figure}
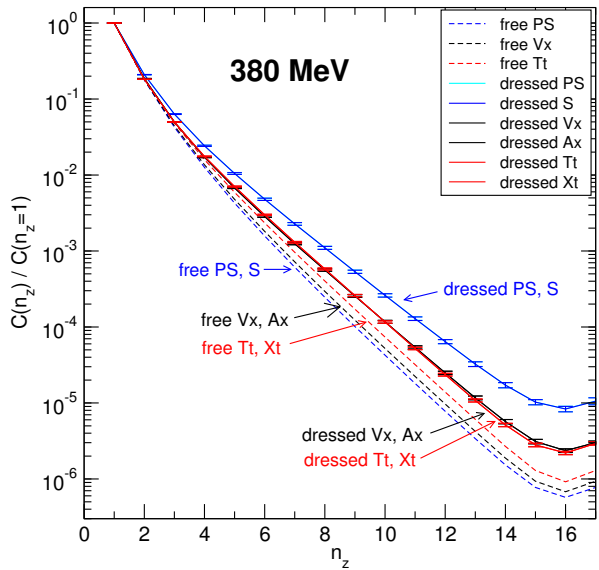

  \centering
  \includegraphics[scale=0.45]{{{2}}}
  \caption{
    $E_1$ and $E_2$ multiplets
    (\protect{\ref{eq:e1}}-\protect{\ref{eq:e2}}) for interacting
    (\textit{dressed}) and non-interacting (\textit{free}) calculations
    at $T$= 380~MeV.
  }
  \label{fig:e2}
\end{figure}

Figure \ref{fig:e2} shows the correlators of the $E_1$ and $E_2$ multiplets
in detail at the highest available temperature $T= 380$~MeV.
Here we also show correlators calculated with non-interacting
quarks. The
non-interacting (\textit{free}) data have been generated on the same
lattice sizes
using a unit gauge configuration
\footnote{Free meson correlators on the lattice have been studied in:
%Stickan:2003gh:
S. Stickan, F. Karsch, E. Laermann and P. Petreczky,
Nucl.Phys.Proc.Suppl. 129 (2004) 599
DOI: 10.1016/S0920-5632(03)02654-9
[arXiv: hep-lat/0309191]}.
Due to the small quark mass the difference
between chiral partners is negligible for the free case, therefore they are
omitted.

We observe a precise degeneracy between $S$ and $PS$ correlators, 
which is consistent with the effective $U(1)_A$ restoration on these 
lattice ensembles \cite{Cossu:2013uua}.
The logarithmic slope of the interacting (\textit{dressed}) \textit{S}
and \textit{PS} correlators is substantially smaller
than that for free quarks. In the latter case the slope is given
by $2\pi/N_t$. A system of two free quarks cannot have `energy' smaller than
twice the lowest Matsubara frequency \cite{DeTar:1987xb}.
For the $E_2$ multiplet we observe asymptotic slopes that are quite
close to $2\pi/N_t$ in agreement with previous studies \cite{Karsch:2003jg}.

\begin{figure}
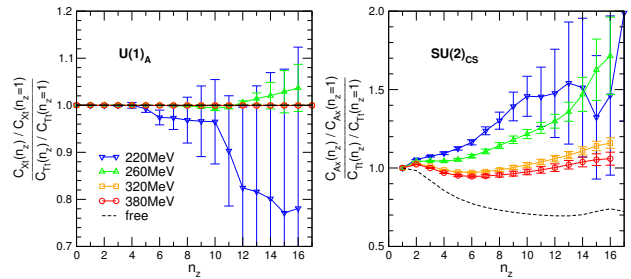

  \centering
  \includegraphics[scale=0.29]{{{4a}}}
  \includegraphics[scale=0.29]{{{4b}}}
  \caption{
    Ratios of normalized correlators from Figure \protect{\ref{fig:corrs}},
    that are related by $U(1)_A$ and $SU(2)_{CS}$ symmetry.
  }
  \label{fig:ratios}
\end{figure}

Figure \ref{fig:ratios} shows
normalized ratios of
$X_t$ and $T_t$ correlators on the left,
as well as of $A_x$ and $T_t$ correlators on the right side.
The $U(1)_A$ symmetry is restored, as is evident from
the left side of this Figure,
where a ratio of correlators in the $X_t$ and $T_t$ 
channels is plotted. We also find a similar degeneracy between $V_x$ and
$A_x$ due to $SU(2)_A$
(See also, {\it e.g.} \cite{Cheng:2010fe,Ishikawa:2017nwl}).

Figures \ref{fig:e2} and \ref{fig:ratios} suggest a possible higher
symmetry ($SU(2)_{CS}$ symmetry, 
see next section) that connects $A_x$ and $T_t$ channels. The right 
panel of Figure 3 shows the corresponding ratio, which demonstrates an 
approximate degeneracy at the level of 5\% above $T\simeq$ 320~MeV. We 
notice that this degeneracy is not expected in the free quark limit 
which is plotted by a dashed curve.
This unexpected symmetry requires that the cross-correlator calculated with the
$A_x$ and $T_t$ operators should vanish.
We have carefully checked that it indeed vanishes to high accuracy.

\begin{figure}
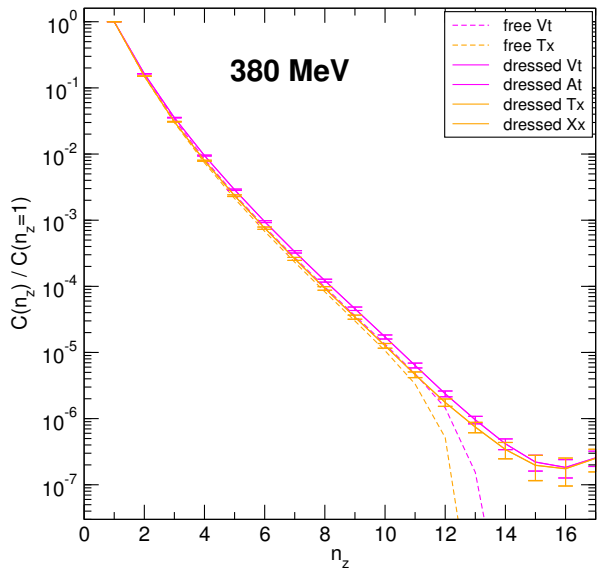

  \centering
  \includegraphics[scale=0.45]{{{3}}}
  \caption{
    $E_3$ multiplet (\protect{\ref{eq:e3}}) for interacting
    (\textit{dressed}) and non-interacting (\textit{free}) calculations.
  }
  \label{fig:e3}
\end{figure}

Figure \ref{fig:e3} shows the $E_3$ multiplet. Here again we observe a precise
degeneracy in all $SU(2)_L \times SU(2)_R$  and $U(1)_A$ connected correlators,
as well as the approximate degeneracy in all four correlators.
We  also see qualitatively
different data between free and dressed correlators at $n_z \geq 11$,
as also seen in \cite{Gavai:2008yv}.
This
implies that we do not observe free non-interacting quarks but  instead
systems with some inter-quark correlation, which is in accordance with
the known results for energy density and pressure at high temperatures
\cite{Karsch:2000ps}.

%%%%%%%%%%%%%%%%%%%%%%%%%%%%%%%%%%%%%%%%%%%%%%%%%%%%%%%%%%%%%%%%%%%%%%%%%%%%%%
%%%%%%%%%%%%%%%%%%%%%%%%%%%%%%%%%%%%%%%%%%%%%%%%%%%%%%%%%%%%%%%%%%%%%%%%%%%%%%
%%%%%%%%%%%%%%%%%%%%%%%%%%%%%%%%%%%%%%%%%%%%%%%%%%%%%%%%%%%%%%%%%%%%%%%%%%%%%%
\section{\label{sec:su4}$SU(2)_{CS}$ and $SU(4)$ symmetries}

In this section we introduce the $SU(2)_{CS}$ and $SU(4)$ transformations,
which connect operators from multiplet $E_2$ (\ref{eq:e2})
as well as from multiplet $E_3$ (\ref{eq:e3})
and contain chiral transformations as a subgroup.
The basic ideas of $SU(2)_{CS}$ and $SU(4)$ symmetries at zero temperature
are given in \cite{Glozman:2015qva}. Here we adapt the group
structure to our setup.

We use the $\gamma$-matrices given by
\begin{equation}
\gamma_i\gamma_j + \gamma_j \gamma_i =
2\delta^{ij}; \qquad \gamma_5 = \gamma_1\gamma_2\gamma_3\gamma_4.
\label{eq:diracalgebra}
\end{equation}
The transformations and generators of the $SU(2)_{CS}$
chiral-spin group,
defined in the Dirac spinor space and diagonal in flavour space, are given by
\begin{equation}
q \rightarrow \exp (\frac{i}{2}\vec \epsilon \cdot \vec \Sigma) q, \quad \vec \Sigma = \{\gamma_k,-i \gamma_5\gamma_k,\gamma_5\}.
\label{eq:su2cs_}
\end{equation}
$SU(2)_{CS}$ contains $U(1)_A$ as a subgroup.
The $\mathfrak{su}(2)$ algebra
$[\Sigma_\alpha,\Sigma_\beta]=2i\epsilon^{\alpha\beta\gamma}\Sigma_\gamma$
is satisfied with any $k=1,2,3,4$. The  $SU(2)_{CS}$ transformations mix the
left- and right-handed components of the quark field. It is not a symmetry
of the free massless quark Lagrangian.
For $z$-direction correlators the following representations
of $SU(2)_{CS}$ are relevant:
\begin{eqnarray}
R_1:\;
\{\gamma_1,-i\gamma_5\gamma_1,\gamma_5\}=
\{\sigma^{23}~ i\gamma_5\gamma_4,\sigma^{23}~\gamma_4,\gamma_5\}, \label{eq:R1} \\
R_2:\;
\{\gamma_2,-i\gamma_5\gamma_2,\gamma_5\}=
\{\sigma^{31}~ i\gamma_5\gamma_4,\sigma^{31}~\gamma_4,\gamma_5\}. \label{eq:R2}
\end{eqnarray}
Those differ from the representation
$\{\gamma_4,-i\gamma_5 \gamma_4,\gamma_5\}$ relevant
for $t$-direction correlators \cite{Glozman:2015qva} by the
rotations
$\sigma^{23} = \frac{i}{2}[\gamma_2,\gamma_3]$
and $\sigma^{31}=\frac{i}{2}[\gamma_3,\gamma_1]$.

These $R_1$ and $R_2$ $SU(2)_{CS}$ transformations connect the following
operators from the $E_2$ multiplet:
\begin{eqnarray}
R_1: &\qquad
A_y \leftrightarrow T_t \leftrightarrow X_t, 
\\
R_2: &\qquad
A_x \leftrightarrow T_t \leftrightarrow X_t,
\end{eqnarray}
as well as  the operators from the $E_3$ multiplet:
\begin{eqnarray}
R_1: &\qquad
A_t \leftrightarrow T_y \leftrightarrow X_y,
\\
R_2: &\qquad
A_t \leftrightarrow T_x \leftrightarrow X_x.
\end{eqnarray}

Our lattice symmetry group includes both the permutation operator 
$\hat P_{xy}$ and $\mathds{1}$ transformations, which form a group
$S_2$.
$\hat P_{xy}$ permutes $\gamma_1$ and $\gamma_2$, and transforms
$\gamma_5$ to $-\gamma_5$.
Then $P_{xy} R_1$ is isomorphic to $R_2$. 
This means that $S_2 \times SU(2)_{CS}$ contains multiplets
\begin{align}
(A_x,A_y,T_t,X_t); \quad (A_t,T_x,T_y,X_x,X_y).
\end{align}

The degeneracy between $\mathbf{V}$ and $\mathbf{A}$ means 
$SU(2)_L \times SU(2)_R$ symmetry. A minimal group that includes 
$SU(2)_L \times SU(2)_R$ and $SU(2)_{CS}$ is $SU(4)$.
The $15$ generators of $SU(4)$ are the following matrices:
\begin{align}
\{
(\tau_a \otimes \mathds{1}_D),
(\mathds{1}_F \otimes \Sigma_i),
(\tau_a \otimes \Sigma_i)
\}
\end{align}
with flavour index $a=1,2,3$ and $SU(2)_{CS}$ index $i=1,2,3$.
Predictions of $S_2 \times SU(4)$ symmetry for isovector operators are the
following multiplets:
\begin{align}
(V_x,V_y,T_t,X_t,A_x,A_y); \; (V_t,T_x,T_y,X_x,X_y,A_t).
\label{eq:su4multiplet}
\end{align}
$S_2 \times SU(4)$ multiplets include in addition the isoscalar partners of
$A_x$, $A_y$, $T_t$ and $X_t$ operators for the first multiplet in
(\ref{eq:su4multiplet})
as well as of $A_t$, $T_x$, $T_y$, $X_x$,$X_y$ for the second multiplet in
(\ref{eq:su4multiplet}).

% cro: i have no idea why, but there is little to no space between
%      Sec 4 & 5, so insert a little manually:
%\vspace{1cm}

%%%%%%%%%%%%%%%%%%%%%%%%%%%%%%%%%%%%%%%%%%%%%%%%%%%%%%%%%%%%%%%%%%%%%%%%%%%%%%
%%%%%%%%%%%%%%%%%%%%%%%%%%%%%%%%%%%%%%%%%%%%%%%%%%%%%%%%%%%%%%%%%%%%%%%%%%%%%%
%%%%%%%%%%%%%%%%%%%%%%%%%%%%%%%%%%%%%%%%%%%%%%%%%%%%%%%%%%%%%%%%%%%%%%%%%%%%%%
\section{\label{sec:conclusions}Conclusions and Discussion}

Our lattice results are consistent with emergence of
approximate $SU(2)_{CS}$ and $SU(4)$ symmetries in 
spin $J=1$ correlators by increasing temperature.
%The data show that the correlation functions
%do not approach the free quark limit.
The considered correlation functions do not seem to approach the
free quark limit.

How could these approximate $SU(2)_{CS}$ and $SU(4)$ symmetries arise
at high temperatures?
They are not symmetries of the QCD Lagrangian.
They are both symmetries of the confining chromo-electric interaction in
Minkowski space since any unitary transformation leaves
%$\bar q \gamma^0 D_0 q = q^\dagger D_0 q$ invariant.
the temporal part of the fermion Lagrangian $\bar{q}\gamma^\mu D_\mu q$ invariant.
The chromo-magnetic interaction 
%contained in $\bar q \vec \gamma D q$
described by the rest of the Lagrangian
breaks both symmetries \cite{Glozman:2015qva}. Consequently the emergence
of these symmetries suggests that the chromo-magnetic interaction is
suppressed at high temperature while the chromo-electric interaction
is still active. This could have implications on the nature of the
effective degrees of freedom in the high temperature phase of QCD since
these symmetries are incompatible with asymptotically free deconfined quarks.

\begin{acknowledgments}
We thank C. Gattringer for numerous discussions.
Support from the Austrian Science Fund (FWF) through the grants
DK W1203-N16 and P26627-N27 is acknowledged.
Numerical calculations are performed on Blue Gene/Q at KEK under its 
Large Scale Simulation Program (No. 16/17-14). This work is supported in 
part by JSPS KAKENHI Grant Number JP26247043 and by the Post-K 
supercomputer project through the Joint Institute for Computational 
Fundamental Science (JICFuS).
G.C. is supported by STFC, grant ST/L000458/1.
S.P. acknowledges the financial support from the Slovenian Research
Agency ARRS (research core funding No. P1-0035).
\end{acknowledgments}

%%%%%%%%%%%%%%%%%%%%%%%%%%%%%%%%%%%%%%%%%%%%%%%%%%%%%%
\section{ Appendix} 
%%%%%%%%%%%%%%%%%%%%%%%%%%%%%%%%%%%%%%%%%%%%%%%%%%%%%%
The symmetry of the $(x,y,t)$--volume (the fixed $n_z$ subvolume, where the
discussed operators are defined) is $D_{4h}$ \cite{grouptheorytext}.
Consider the  transformations of the Euclidean interpolators
$O(n)=\bar q(n) \Gamma q(n)$ with $n=(n_x,n_y,n_z,n_t)$:
\begin{align}
%\mathrm{rot}&: \;\;
%\begin{aligned}
%\bar q(n) \Gamma& q(n) \to \\
%\bar q(n) &\exp(\tfrac{i}{2} \omega_{\mu\nu} \sigma^{\mu\nu}) \Gamma \exp(-\tfrac{i}{2} \omega_{\mu\nu} \sigma^{\mu\nu})q(n)\nonumber\\
%\end{aligned} \\[8pt]
%\hat P^z &: \;\;
%\begin{aligned}
%\bar q(n) \Gamma q(n)& \to \\ 
%\bar q( n_{P^z})& \gamma_3 \Gamma \gamma_3 q( n_{P^z})
%\end{aligned}
\mathrm{rot}&: \quad
\bar q \Gamma q \to
\bar q \exp(\tfrac{i}{2} \omega_{\mu\nu} \sigma^{\mu\nu}) \Gamma \exp(-\tfrac{i}{2} \omega_{\mu\nu} \sigma^{\mu\nu})q\nonumber\\
\hat P^z&: \quad
\bar q \Gamma q \to 
\bar q \gamma_3 \Gamma \gamma_3 q
\end{align}
under discrete rotations and $\hat P^z$ that performs
inversion $n \to n_{P^z}=(-n_x,-n_y,n_z,-n_t)$.    
 
The relevant symmetry group $D_{4h}$ has  ten classes of group elements
and ten irreducible representations identified by characters in Table
\ref{tab:characters}:
$C_4$ and $C_2$ are rotations around $t$ for $\pi/2$ and $\pi$,
respectively, $C_2^\prime$ is a rotation for $\pi$ around $x$,
while $C_2^{\prime\prime}$ is a rotation for $\pi$ around $x+y$.
Further five classes are obtained by multiplication of the
elements with $\hat P^z$ and the characters with  $(-1)^{P^z}$. 
  
According to these transformations, the interpolators
for $\mathbf{V,A,T,X}$ operators of Table \ref{tab:ops} transform
under the irreducible representations given in
Table \ref{tab:transformations}. Note that group elements
of $D_{4h}$ transform interpolators only within one box of
that table and indeed the observed energy levels for the
$x$ and $y$ components of an operator agree. 
However, no group element of $D_{4h}$ transforms components
of $\mathbf{V}$ to $\mathbf{T}$ (or $\mathbf{A}$ to $\mathbf{X}$). 

\begin{table}[htb]
\begin{tabular}{c|ccccc}
  \hline \hline
  & $\mathrm{Id}$ & $C_4$ & $C_2$ & $C_2^\prime$ & $C_2^{\prime \prime}$  \\
  \hline
  $A_1^\pm $ & 1 & 1 & 1 & 1 & 1 \\ 
  $A_2^\pm $ & 1 & 1 & 1 & -1 & -1 \\ 
  $B_1^\pm $ & 1 & -1 & 1 & 1 & -1 \\
  $B_2^\pm $ & 1 & -1 & 1 & -1 & 1 \\
  $E^\pm $ & 2 & 0 & -2 & 0 & 0 \\
\hline \hline \end{tabular}
\caption{The first five columns of the character table
         of $D_{4h}$ \cite{grouptheorytext}; the superscript
         in the irreps denotes  $P_z$. The first column
         gives the dimension of the irrep. Another five
         classes are obtained by multiplication of elements
         with $P^z$, while the characters get a factor of
         $(-1)^{P^z}$. 
}\label{tab:characters}
\end{table}

\begin{table}[htb]
\begin{tabular}{| c  | c  | c  | c | c | c |} 
  \hline
   $V_x,V_y:\ E^- $ & $ V_z: \ A_1^+$ & $  V_t:\ A_2^-$   \\ 
   \hline 
   $A_x,A_y:\ E^+ $ & $ A_z: \ A_1^-$ & $  A_t:\ A_2^+$   \\  
   \hline
   $T_x,T_y:\ E^- $ & $ T_z: \ A_1^+$ & $  T_t:\ A_2^-$   \\  
   \hline
   $X_x,X_y:\ E^+ $ & $ X_z: \ A_1^-$ & $  X_t:\ A_2^+$   \\  
  \hline \end{tabular}
\caption{The interpolators $O(n)$, which are denoted by
         $\mathbf{V,A,T,X}$ in Table \ref{tab:ops},
         transform according to the irreps in this Table.}
\label{tab:transformations}
\end{table}

%%%%%%%%%%%%%%%%%%%%%%%%%%%%%%%%%%%%%%%%%%%%%%%%%%%%%%%%%%%%%%%%%%%%%%%%%%%%%%
%%%%%%%%%%%%%%%%%%%%%%%%%%%%%%%%%%%%%%%%%%%%%%%%%%%%%%%%%%%%%%%%%%%%%%%%%%%%%%
%%%%%%%%%%%%%%%%%%%%%%%%%%%%%%%%%%%%%%%%%%%%%%%%%%%%%%%%%%%%%%%%%%%%%%%%%%%%%%

\end{document}